# Cesium-involved electron transfer and electron-electron interaction in high-pressure metallic CsPbI$_3$

Feng Ke,[1,2] Jiejuan Yan,[2] Shanyuan Niu,[1,2] Jiajia Wen,[1] Ketao Yin,[3]* Nathan R. Wolf,[4] Yan-Kai Tzeng,[5] Hemamala I. Karunadasa,[1,4] Young S. Lee,[1,6] Wendy L. Mao,[1,2] Yu Lin[1]*

[1]*Stanford Institute for Materials and Energy Sciences, SLAC National Accelerator Laboratory, Menlo Park, California 94025, United States*

[2]*Department of Geological Sciences, Stanford University, Stanford, California 94305, United States*

[3]*School of Physics and Electronic Engineering, Linyi University, Linyi, Shandong 276005, China*

[4]*Department of Chemistry, Stanford University, Stanford, California 94305, United States*

[5]*Department of Physics, Stanford University, Stanford, California 94305, United States*

[6]*Department of Applied Physics, Stanford University, Stanford, California 94305, United States*

*To whom correspondence should be addressed. E-mail: yinketao@lyu.edu.cn; lyforest@stanford.edu

**Abstract**

Electron-phonon coupling was believed to govern the carrier transport in halide perovskites and related phases. Here we demonstrate that electron-electron interaction plays a direct and prominent role in the low-temperature electrical transport of compressed $CsPbI_3$ and renders Fermi liquid (FL)-like behavior. By compressing δ-$CsPbI_3$ to 80 GPa, an insulator-to-metal transition occurs, concomitant with the completion of a sluggish structural transition from the one-dimensional (1D) *Pnma* (δ) phase to a 3D *Pmn*$2_1$ (ε) phase. Deviation from FL behavior is observed in $CsPbI_3$ upon entering the metallic ε phase, which progressively evolves into a FL-like state at 186 GPa. First-principles density functional theory calculations reveal that the enhanced electron-electron coupling is related to the Cs-involved electron transfer and sudden increase of the 5*d* state occupation of the high-pressure ε phase. Our study presents a promising strategy for tuning the electronic interaction in halide perovskites for realizing intriguing electronic states.

3D halide perovskites of the form $ABX_3$ (A = organic or inorganic cation, B = metal cation, and X = halide anion) display remarkable optical and electronic properties that find applications for a wide range of technologies such as photovoltaics[1,2] and light-emitting diodes[3]. In these materials, $BX_6$ octahedra corner-share to form the 3D perovskite structure. Beyond the perovskite phase, $BX_6$ octahedra can also connect through edge- and face-sharing and extend into related 2D and 1D non-perovskite structures. At ambient conditions, the electronic states of the metal-halide sublattice are typically responsible for the valence band maximum (VBM) and conduction band minimum (CBM) of these $ABX_3$ phases, and govern their optical absorption and electrical transport properties[4-6]. It is believed that the A-site cation in $ABX_3$ phases only makes a minimal contribution to the band edge states and plays an indirect role in the electronic structures of these materials[7,8].

Modulating the electronic structures and searching for novel electronic states in $ABX_3$ phases have attracted considerable experimental and theoretical attention, partly motivated by their structural similarity to the perovskite oxides where exciting electronic states such as ferroelectricity[9], unconventional superconductivity[10,11], and topological insulating behavior[12] have been discovered. Recent experimental studies observed ferroelectric-like behavior in $(MA)PbI_3$ (MA = $CH_3NH_3^+$) at ambient pressure, although this topic is still under debate[13-15]. Compositional tuning or designing lattice architectures are effective methods for modifying their electronic structures and carrier transport properties[16]. Lattice compression has also been found to be a useful means for accessing different electronic states[17-25]. Metallic-like behavior was experimentally

achieved in compressed (MA)PbI$_3$, (FA)PbI$_3$ (FA = CH(NH$_2$)$_2$$^+$), and δ-CsPbI$_3$ [26-28]. Bandgap closure and reopening were observed in CsAuI$_3$ at high pressure[29]. However, almost all these high-pressure experimental efforts were conducted near room temperature. Theoretically, it is predicted that under compression halide perovskites could be topological insulators – *e.g.,* cubic CsPbI$_3$ [30-32], or superconductors – *e.g.,* CsTlX$_3$ [33].

At low temperatures, correlated materials can display fascinating phenomena such as various types of metal-insulator and superconducting transitions due to strong electron-electron (*e-e*), electron-phonon (*e-p*), and electron-magnon interaction. The carrier transport and scattering mechanisms have well known temperature dependences. For instance, in simple metals the *e-p* interaction plays an important role in the carrier transport, leading to a $T^5$ dependence of resistivity at low temperatures with a linear temperature dependence above the Debye temperature[34]. For interacting fermion systems, the carrier scattering has been well described by the Fermi liquid (FL) model[35], in which carriers are considered as quasiparticles and the *e-e* interaction dominates the low-temperature electrical transport, causing a $T^n$ ($n = 2$) dependence of resistivity for 3D metals. In certain correlated materials, FL behavior has been observed to break down, resulting in non-Fermi liquid (NFL) behavior with $n < 2$, *e.g.,* in the normal state of high-temperature iron- and cuprate-based superconductors[36-39]. ABX$_3$ phases were reported to have a Debye temperature of 100 – 250 K[40,41]. Hence carrier transport behavior at low temperature will be different from that at above the Debye temperature. At ambient pressure, *e-p* coupling was believed to be mainly responsible for the



scattering of the free charge carriers in ABX$_3$-based materials[42-44], although a recent study indicated that *e-e* interaction had contributions to the optical properties, such as increasing the absorption coefficient[45]. Studying the electronic structures and carrier scattering mechanisms in these materials at low temperature combined with high pressure conditions represents unexplored territory.

In this work, we study novel electronic states in CsPbI$_3$ over a vast pressure-temperature space of 0.1 – 186 GPa and 2 – 300 K. With application of pressure to 80 GPa, insulating δ-CsPbI$_3$ transforms to a metallic state, accompanied by the completion of a sluggish *Pnma*-to-*Pmn*2$_1$ (δ-to-ε) phase transition from a 1D chain structure to a 3D structure. Deviation from a FL-like state is observed as CsPbI$_3$ enters the metallic ε phase. With further compression to 186 GPa, the deviation gradually disappears, and the metallic phase behaves as a Fermi liquid. Band structure and electron transfer calculations reveal dramatically enhanced electron transfer and sudden increase of the 5*d* state occupation of Cs and I in the ε phase that strengthen the *e-e* interaction and induce the FL-like behavior. In contrast to ambient conditions where the A-site cation only has an indirect effect on the electronic properties of the ABX$_3$ phases, the Cs cation is found to play a direct and prominent role in the electronic structure of CsPbI$_3$ at high pressure.

**Results**

We measured the resistivity of CsPbI$_3$ over a temperature range of 2 – 300 K and pressures up to 186 GPa (Figs. 1 and 2). Sample characterization and experimental



details can be found in the supporting information (Supplementary Figs. 1-4). At ambient conditions, the room-temperature resistivity ($\rho_{RT}$) is beyond the measurable range of the instrument since δ-CsPbI$_3$ is a large bandgap (2.5 – 2.8 eV) insulator[46,47]. With increasing pressure, the value of $\rho_{RT}$ falls into the measurable range and is 4.5 × 10$^3$ Ω·cm at 21.5 GPa, accompanied by the color of the sample changing from yellow to red and then to black. With further compression to 80 GPa, $\rho_{RT}$ drops dramatically by more than 6 orders of magnitude, beyond which a linear reduction of Log $\rho_{RT}$ at a rate of (5.3±0.4) × 10$^{-3}$ Ω·cm/GPa continues up to 186 GPa, suggesting an electronic transition at 80 GPa.

The resistivity-temperature ($\rho$-$T$) curves of CsPbI$_3$ at representative pressures are shown in Fig. 2. At pressures below 50 GPa, the temperature dependence of resistivity (d$\rho$/d$T$) was negative throughout the temperature range measured (2 – 300 K), indicating an insulating character. By fitting the linear range of the ln $\rho$ – 1/$T$ curves at high temperature to the Arrhenius equation, an activation energy of 134.8 ± 5.3, 59.1 ± 2.1, and 29.5 ± 0.7 meV is obtained at 30, 41, and 50 GPa, respectively. Upon further compression to 61 GPa, the resistivity starts to decrease with cooling (d$\rho$/d$T$ > 0, metallic behavior) near room temperature but reverses its temperature dependence back to d$\rho$/d$T$ < 0 below ~255 K. The transition temperature shifts to ~135 K at 70 GPa. A linear extrapolation suggests that at a critical pressure of 80.5 GPa, the sign reversal of d$\rho$/d$T$ at low temperature could be fully suppressed. The experimental result at 80 GPa indeed shows a positive d$\rho$/d$T$ throughout the temperature range of 2 – 300 K, indicating the electronic transition to a metallic state in compressed CsPbI$_3$.



Detailed analysis of the $\rho$-$T$ curves reveals remarkable behavior in the metallic phase. For $T > 110$ K, a linear temperature dependence is obtained in all the $\rho$-$T$ curves above 80 GPa (Fig. 2a), which is consistent with the well-known simple metal model where the $e$-$p$ interaction dominates the carrier transport. The $\rho$-$T$ relationship obeys the following equation,

$$\rho(T) = \rho_0 + BT^5 + CT \qquad (1)$$

where $\rho_0$ is the residual resistivity, and B and C are the coefficients of $T^5$ and $T$-terms, respectively. Surprisingly, the $\rho$-$T$ curves below 110 K significantly deviate from this $T$-dependence (Fig. 2b, and Supplementary Figs. 5 and 6). Instead, $\rho(T)$ is found to be proportional to $T^n$,

$$\rho(T) = \rho_0 + AT^n \qquad (2)$$

where the fitted exponent $n$ is $1.75 \pm 0.02$ at 80 GPa and increases gradually with pressure to $1.98 \pm 0.03$ at 186 GPa (Fig. 2c). The observation of $\rho(T) \sim T^n$ where $n \leq 2$ in compressed CsPbI$_3$ at low temperature is anomalous. Such behavior has typically been observed in materials such as high-temperature superconductors and heavy fermion compounds[36-39], in which the $d$ or $f$ electron shell is partially filled and $e$-$e$ interaction is important. However, at ambient conditions δ-CsPbI$_3$ does not have any $d$-electrons that can readily serve as charge carriers.

To understand the abnormal low-temperature behavior of the high-pressure metallic phase, we studied the structural evolution of δ-CsPbI$_3$ as a function of pressure using powder X-ray diffraction (XRD) measurements (Fig. 3a and Supplementary Fig. 7). Below 5.4 GPa, δ-CsPbI$_3$ crystallizes in a 1D double chain structure with a *Pnma* symmetry where PbI$_6$ octahedra edge-share along the *b*-axis (Supplementary Fig. 9), consistent with the reported structure at ambient conditions[28,48]. Upon compressing to 6.7 GPa, new diffraction peaks (marked as red arrows in Fig. 3a and Supplementary Fig. 7) start to appear and more peaks develop with further compression, indicating the emergence of a high-pressure phase (ε-CsPbI$_3$). This δ-to-ε structural transition is found to be very sluggish where the high-pressure phase grows at the expense of the low-pressure phase. For instance, the peak that appears at ~9.8° (red triangle in Fig. 3a) at 6.7 GPa rapidly increases its intensity and becomes the strongest peak above 20.3 GPa, accompanied by a dramatic intensity decrease of the (211) and (212) diffraction peaks of the low-pressure phase. The (211) and (212) reflections are the most intense peaks of the starting structure and eventually disappear by 82 GPa, indicating the completion of the phase transition that spans a pressure window of ~75 GPa. The XRD patterns above 82 GPa are similar except for peak shifts to higher diffraction angles and subtle changes in the peak widths and intensities. It is noticed that, between 6.7 – 82 GPa, new reflections that can be indexed into the high-pressure ε phase gradually appear upon compression, and the relative peak intensities of the δ phase change with pressure. These observations indicate that the structures of the δ and ε phases evolve within this pressure range. Rietveld refinement indicates that an orthorhombic structure with a *Pmn*2$_1$ space group fits the diffraction patterns above 82 GPa well (Supplementary Fig. 10). In contrast to the initial 1D double chain structure, the high-pressure ε phase is a 3D structure with adjacent double chains bonding together where the Pb atoms are eight- and nine-fold coordinated by the I atoms (Fig. 3c and Supplementary Fig. 9).

Comparing the electrical transport and XRD results, we find that the insulator-to-metal transition and the *Pnma*-to-*Pmn*$2_1$ phase transition are closely related. The pressure at which the structural transition completes (~82 GPa) coincides with the end pressure of the insulator-to-metal transition (~80 GPa) and with that of the change in the $\rho_{RT}$ – $P$ curve. Further analysis of the lattice parameters of the *Pmn*$2_1$ phase show pronounced deviations between 62 – 82 GPa (Supplementary Fig. 11), especially for the *a* lattice constant. This pressure window is consistent with that of the insulator-to-metal transition which starts at ~61 GPa where CsPbI$_3$ shows a metallic character above ~255 K and completes at ~80 GPa where a metallic state is present throughout 2 – 300 K temperature range.

First-principles density functional theory (DFT) calculations corroborate the experimental results (Fig. 3b). We performed extensive structural searches on CsPbI$_3$ using the CALYPSO software package, which can predict energetically stable structures of materials at extreme conditions, *e.g.,* high pressure[49]. At 90 GPa, the structural searches find that an orthorhombic *Pmn*$2_1$ phase and two monoclinic structures with *Cm* and *C*2/*m* symmetries are energetically stable and have lower enthalpies than the starting *Pnma* δ-CsPbI$_3$. Among these structures, only the *Pmn*$2_1$ structure produces a simulated XRD pattern that is consistent with the experimental pattern (Supplementary Fig. 10). After structure optimization, the detailed structural parameters of the *Pmn*$2_1$ phase agree well with those obtained from XRD refinement (Supplementary Table 1). The difference in the transition pressure between experiments (6.7 GPa, Fig. 3a and Supplementary Fig. 7) and calculations (30 GPa, Fig. 3b) may be due to the nature of the very sluggish phase transition that experimentally begins at 6.7 GPa and completes by ~82 GPa. In addition, temperature differences between the XRD experiments (~298 K) and structural searches (0 K) would in part explain the different transition pressures. Our theoretical results indicate that the δ-to-ε structural evolution involves a sequence of Pb-I bond breaking in the starting $PbI_6$ octahedral chains, and formation of Pb-I bonds between adjacent chains under compression (Fig. 3d and Supplementary Fig. 14), consistent with the XRD results that suggest the system is not a simple mixing of the δ and ε phases. It would rather be visualized as a set of mixed states composed of imperfect, distorted structures of δ-$CsPbI_3$ and ε-$CsPbI_3$ that keep evolving with pressure. Our calculations are unable to factor in the complex nature of the mixed phases, which further explains the underestimation of the simulated pressure for the structural and electronic transitions.

Band structure and density of states calculations were performed to study the electronic structure of the high-pressure phase. Because the low-temperature electrical transport results suggest the possibility of *d*-electrons serving as the charge carriers, the 5*d*-states of the Cs and I atoms were also included in our calculations which are close to the Cs-6*s* and I-5*p* states. At 0 GPa of δ-$CsPbI_3$, the VBM is mainly composed of the I-5*p* and Pb-6*s* states, while the Pb-6*p* and I-5*p* states contribute to the CBM (Fig. 4a and Supplementary Fig. 17). The A-site Cs atom makes very little contribution to the VBM or CBM, consistent with previous results[6,28]. The involvement of *d*-states is negligible except for a very small contribution from the I-5*d* state to the upper conduction band at ~3 – 4 eV above the CBM. An obvious transition occurs at high pressure as the material undergoes the δ-to-ε phase transition. In the $Pmn2_1$ ε phase, the electronic states delocalize dramatically, and ε-CsPbI$_3$ has a zero bandgap at 30 GPa (Fig. 4b and Supplementary Fig. 18). If not considering the phase transition, our calculations indicate that δ-CsPbI$_3$ would close the bandgap at 40 GPa. These results further confirm the correlated nature of the structural and electronic transitions in compressed CsPbI$_3$. After examining the band structures of ε-CsPbI$_3$, we found that the A-site Cs cation significantly affects the electronic structure. The Cs-5$d$ and I-5$d$ states contribute appreciably to the valence and conduction bands, which become more pronounced with pressure up to 180 GPa (dashed lines, Fig. 4b), indicating pressure-induced electron transfer. Quantitative electron transfer calculations also support the pressure-induced electron transfer. With the application of pressure, the 5$d$ state occupation of the Cs and I atoms gradually increases, and then a sudden rise is clearly observed when the insulating δ phase transforms into the metallic ε phase (Fig. 4c). Similar electron transfer has also been observed in the elemental form of Cs and I$_2$ [50].

**Discussion**

The pressure-induced enhancement of the 5$d$ state occupation explains the $\rho(T) \sim T^n$ ($n \approx 2$) behavior in compressed CsPbI$_3$. Similar to material systems such as high-temperature superconductors and heavy fermion compounds where $d$ or $f$ electrons cause strong $e$-$e$ interaction[36-39], the 5$d$-electrons in compressed CsPbI$_3$ enhance the $e$-$e$ interaction and direct the electrical transport at low temperature. The observed $\rho$-$T$ curves of the metallic ε phase deviate from the predictions of a simple metal at low temperature ($\rho(T) \sim T^5$), and show FL-like behavior ($\rho(T) \sim AT^n$, $n \approx 2$), suggesting the presence of strong correlations. The coefficient A, which typically characterizes the strength of the *e-e* interaction and is related to the effective mass[51,52], also suggests strong *e-e* interaction (Fig. 2c). The fitted A value is $1.078 \times 10^{-8}$ Ω·cm/K$^n$ at 80 GPa, comparable with those obtained in high-temperature superconductors[52-54]. With further compression, the A value gradually decreases to $1.073 \times 10^{-9}$ Ω·cm/K$^n$ at 186 GPa.

Deviation from a FL-like state ($n < 2$) is also observed at the onset of CsPbI$_3$ entering the high-pressure metallic phase (Fig. 2c). Many origins for the deviation from FL behavior may be at play in correlated electron systems. Factors such as the Kondo effect[55,56] and lattice deformation and disorder[56-58] have been proposed. NFL behavior has also been observed in some unconventional superconductors and heavy fermion metals that was attributed to quantum critical fluctuations[37,39,51]. The Kondo effect usually occurs in correlated metals doped with magnetic impurities. Magnetic impurities are not known to exist in CsPbI$_3$ (Supplementary Fig. 1), thus ruling out the Kondo effect as being the likely cause. It is possible that the high-pressure metallic CsPbI$_3$ phase has defects/disorder in the lattice. However, extrinsic disorder like grain boundaries usually increases the residual resistivity ($\rho_0$) but does not change the temperature dependence of resistivity; intrinsic disorder like lattice defects usually leads to a negative value for the coefficient A[56,57,59], inconsistent with our electrical transport results (Fig. 2c). The evolution of *n* from $1.75 \pm 0.02$ at 80 GPa to $1.98 \pm 0.03$ at 186 GPa also suggests that pressure-induced defects and disorder are not the root cause of the deviation in the high-pressure metallic phase because they are present throughout the compression process. The electrical transport anomaly in compressed $CsPbI_3$ seems similar in character to some unconventional superconductors and heavy fermion metals, in which NFL behavior has been observed to exist for a wide range of doping levels between the under-doped and over-doped compositions that may be attributed to proximity to a nearby quantum critical point[36-39]. Pressure induces the insulator-to-metal transition and increases the carrier concentration in $CsPbI_3$, which resemble the evolution from an under- to an over-doped composition in unconventional superconductors and heavy fermion metals. This raises the possibility that the deviation from a FL-like state in the high-pressure metallic ε-$CsPbI_3$ is related to quantum criticality. In iron- and cuprate-based superconductors, such behavior usually coexists with a superconducting state in the phase diagram[36,37]. Hence, a superconducting transition might occur below 2 K in the high-pressure metallic $CsPbI_3$. Further experiments are required to understand the origin of this anomaly and search for a superconducting phase in $CsPbI_3$ and related materials.

In summary, by compressing δ-$CsPbI_3$ to 80 GPa, the initially insulating phase transforms to a metallic state, accompanied by the completion of a sluggish δ-to-ε phase transition from a 1D chain structure to a 3D structure that begins at 6.7 GPa and finishes at ~82 GPa. Deviation from a FL-like state is observed in the high-pressure metallic phase, which progressively evolves into a FL-like state at ~186 GPa. First-principles DFT calculations indicate that pressure induces electron transfer which significantly enhances the 5*d* state occupation of Cs and I in the ε phase that strengthens the *e-e* interaction and eventually leads to the formation of a FL-like state. In particular, the Cs atom is found to make a direct and pronounced contribution to the electrical properties of CsPbI$_3$ at high pressure. Our study opens opportunities for realizing novel electronic states and physical properties in halide perovskites through pressure engineering and cationic manipulation.

**Methods**

Yellow δ-CsPbI$_3$ samples were synthesized using the previously reported solution-based method[60]. Resistivity measurements were conducted using a nonmagnetic diamond anvil cell with beveled diamonds with an inner and outer culet diameter of 100/300 μm in a physical property measurement system (PPMS DynaCool). High-pressure XRD experiments were performed at beamline 12.2.2 of the Advanced Light Source, LBNL. Structure prediction was performed via a global minimum search of the free energy surface by the swam intelligence-based CALYPSO method[61]. More experimental and calculation details can be found in Supplementary information.

**Data availability**

All data that support the finding of this study are present in the paper and the supplementary information.

**Acknowledgements**

This work was supported by the Department of Energy (DOE), Office of Science, Basic Energy Sciences, Materials Sciences and Engineering Division (DE-AC02-76SF00515). Beamline 12.2.2 is a DOE Office of Science User Facility under contract no. DE-AC02-05CH11231. Portions of this work were performed at HPCAT (Sector 16), Advanced Photon Source (APS), Argonne National Laboratory (ANL). HPCAT operations are supported by DOE-NNSA's Office of Experimental Sciences. The APS is a DOE Office of Science User Facility operated for the DOE Office of Science by ANL under Contract No. DE-AC02-06CH11357. N.R.W. was partially supported by a Stanford Interdisciplinary Graduate Fellowship. The simulation work by K.Y. was supported by the Natural Science Foundation of China under Grant No. 11904148. We thank Dr. Chunjing Jia and Wen Wang for helpful discussion.

**Author contributions**

F.K. and Y.L. designed the project and wrote the paper. F.K., J.Y., S.N., J.W., Y.S.L. Y.T., W.L.M., and Y.L. conducted the experiments and analyzed the data. N.R.W synthesized the sample under H.I.K.'s supervision. K.Y. performed the calculations. All authors contributed to the discussion and revision of the paper.

F.K., J.Y., and S.N. contributed equally to this work.

The authors declare no competing financial interests.

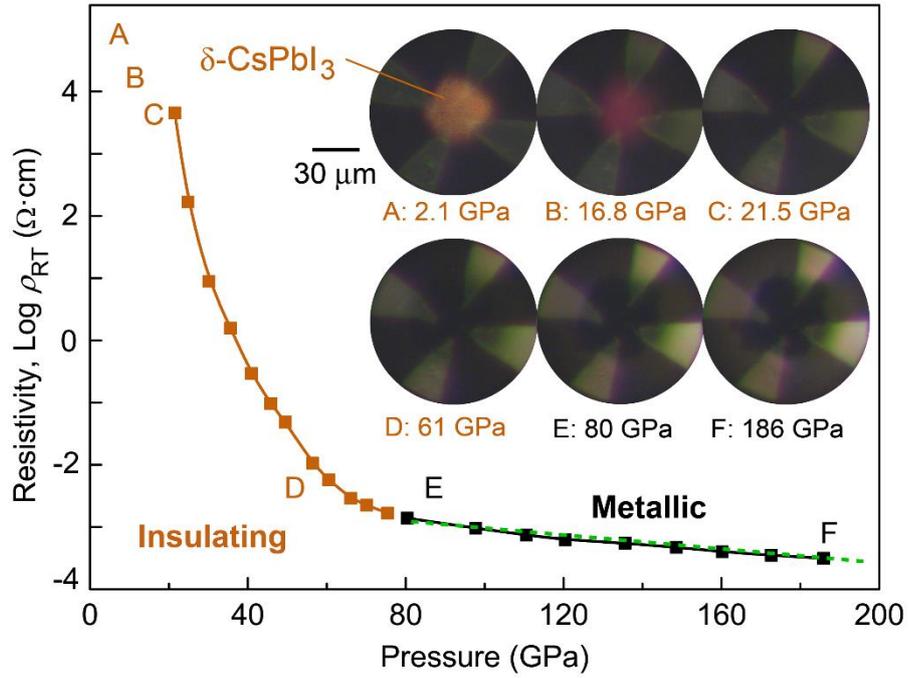

**Fig. 1. Room-temperature resistivity (Log $\rho_{RT}$) of CsPbI$_3$ as a function of pressure.** The green dashed line is a linear fit of the Log $\rho_{RT}$ – P curve above 80 GPa. A slope of -(5.3±0.4) × 10$^{-3}$ Ω·cm/GPa is obtained. The photomicrographs show the hand-wired electrode configuration for the high-pressure resistivity measurements and the color change of the sample from yellow to red and further to opaque black under compression.

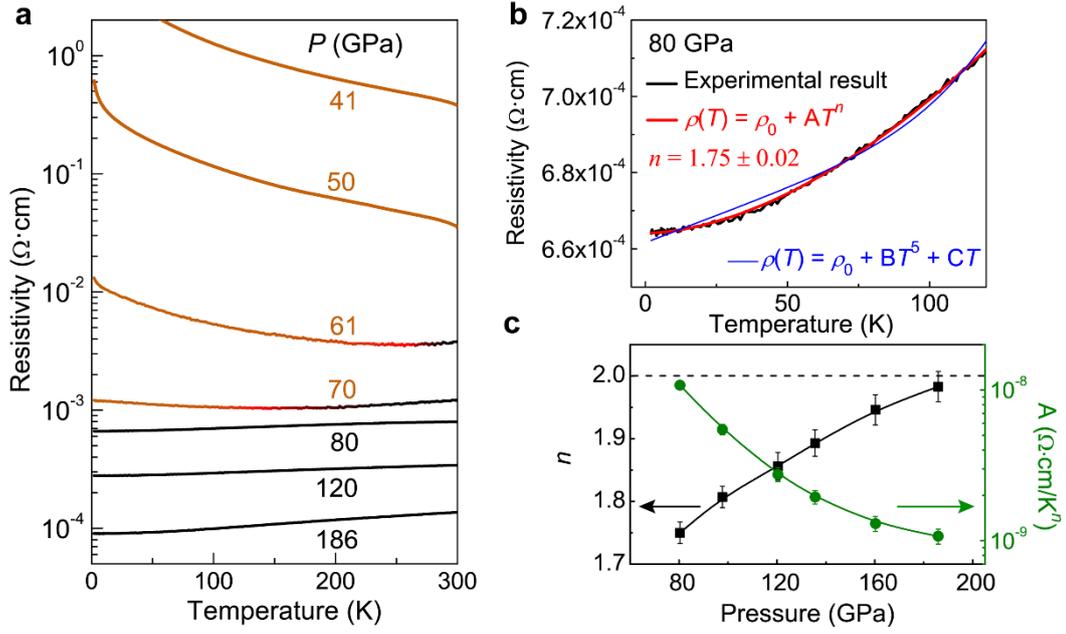

**Fig. 2. Low-temperature electrical transport properties of CsPbI$_3$ at high pressure.**

**a,** Resistivity–temperature ($\rho$–$T$) curves at representative pressures showing the insulating-to-metallic transition in compressed CsPbI$_3$. **b,** Fitting of the $\rho$–$T$ at 80 GPa to the models of $\rho(T) = \rho_0 + AT^n$ (red line) and $\rho(T) = \rho_0 + BT^5 + CT$ (blue line). **c,** The exponent $n$ (black squares) and coefficient A (green circles) obtained using the $\rho(T) = \rho_0 + AT^n$ model as a function of pressure.

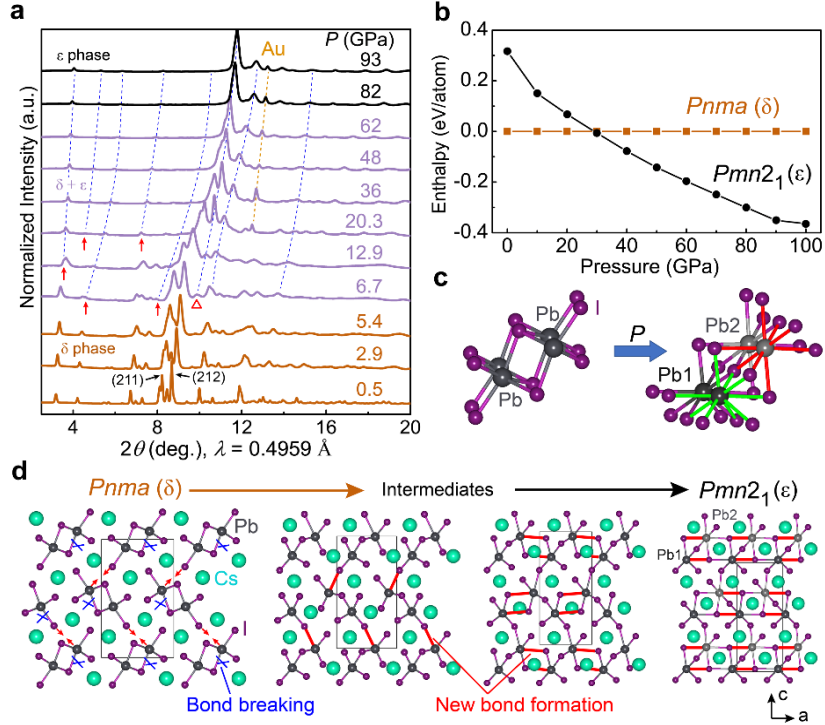

**Fig. 3. Experimental and theoretical evidence for the pressure-induced δ-to-ε structural transition in $CsPbI_3$. a**, Selected XRD patterns of compressed $CsPbI_3$ showing the gradual δ-to-ε phase transition. See Supplementary Fig. 7 for the full set of XRD patterns. The red arrows mark the appearance of new diffraction peaks. The red triangle marks the diffraction peak at 9.8° which rapidly increases in intensity and becomes the strongest peak above 20.3 GPa. The blue dashed lines indicate the evolution of the sample peak positions with pressure. The yellow dashed line tracks a diffraction peak of gold (Au) which is the internal pressure standard. **b**, Calculated enthalpy of the ε phase relative to the δ phase as a function of pressure. **c**, The Pb coordination number change from low to high pressure. The green and red bonds indicate the nine- and eight-fold coordination of Pb1 (dark gray) and Pb2 (light gray) atoms, respectively. **d**, A predicted structural transition path from the initial δ to the high-pressure ε phase.

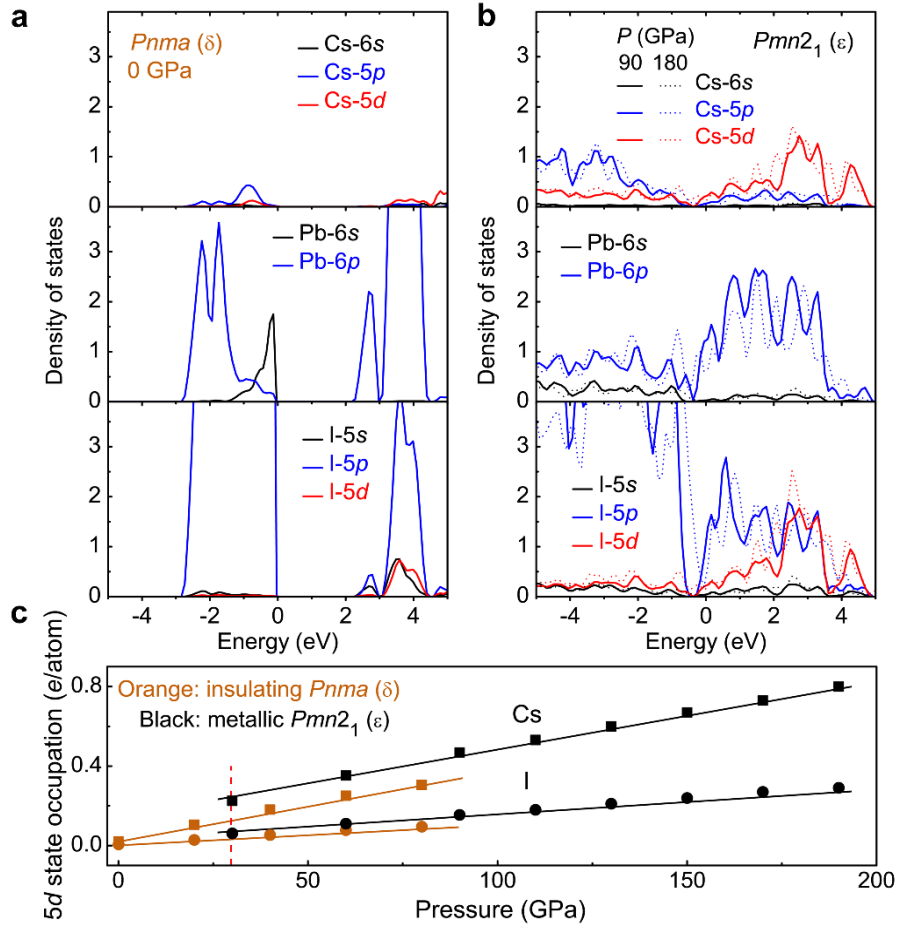

**Fig. 4. The electronic structure of CsPbI₃ at representative pressures. a**, Density of states of the δ phase at 0 GPa. **b**, Density of states of the ε phase at 90 GPa (solid lines) and 180 GPa (dashed lines). **c**, The 5$d$ state occupation of the Cs (squares) and I (circles) atoms in the initial δ (orange) and high-pressure ε (black) phases as a function of pressure. The red vertical dashed line indicates the transition pressure from the enthalpy calculations.

23